\documentclass[11pt]{amsart}
\usepackage[margin=1in,letterpaper]{geometry}

\usepackage{amssymb,amsfonts,amsmath,bbm,mathrsfs,stmaryrd,wasysym}

\usepackage{xcolor}
\usepackage[colorlinks,
             linkcolor=black!75!red,
             citecolor=blue!50!black,
             pdftitle={A combinatorial universal star-product},
             pdfauthor={Theo Johnson-Freyd},
             pdfsubject={quantization},
             pdfkeywords={quantization,diagrams},
             pdfproducer={pdfLaTeX},
             pdfpagemode=None,
             bookmarksopen=true
             bookmarksnumbered=true]{hyperref}

\usepackage{amsthm}

\newtheorem{definition}{Definition}[section]
\newtheorem{proposition}[definition]{Proposition}
\newtheorem{theorem}[definition]{Theorem}
\newtheorem{maintheorem}[definition]{Main Theorem}
\newtheorem{corollary}[definition]{Corollary}

\renewcommand\qedhere{\hfill\qedsymbol}

\usepackage{tikz}
\usetikzlibrary{arrows,decorations.pathmorphing,decorations.pathreplacing,decorations.markings,shapes.geometric,through,fit,shapes.symbols}
\tikzset{
    dot/.style={circle,draw,fill,inner sep=1.25pt},
    intdot/.style={circle,draw,fill=gray,inner sep=2pt},
    opendot/.style={circle,draw,inner sep=1pt},
    onearrow/.style={postaction={decorate}, decoration={markings,mark=at position .6 with {\arrow[draw,line width=1pt]{>}}}},
    inversearrow/.style={postaction={decorate}, decoration={markings,mark=at position .45 with {\arrow[draw,line width=1pt]{<}}}},
    twoarrows/.style={draw, postaction={decorate}, decoration={markings,mark=at position .35 with {\arrow[draw,line width=1pt]{>}},mark=at position .75 with {\arrow[draw,line width=1pt]{>}}}},
    twoarrowsempty/.style={postaction={decorate}, decoration={markings,mark=at position .3 with {\arrow[draw,line width=1pt]{>}},mark=at position .7 with {\arrow[draw,line width=1pt]{>}}}},
    inversetwoarrows/.style={draw, postaction={decorate}, decoration={markings,mark=at position .35 with {\arrow[draw,line width=1pt]{<}},mark=at position .7 with {\arrow[draw,line width=1pt]{<}}}},
    squiggly/.style={draw, decorate,decoration={snake,amplitude=.3mm,segment length=2mm}},
    fastsquiggly/.style={draw, decorate,decoration={snake,amplitude=.3mm,segment length=1mm}},
    inversesquiggly/.style={draw, decorate,decoration={snake,amplitude=.2mm,segment length=2mm},postaction={decorate,decoration={markings,mark=at position .45 with {\arrow[draw,line width=1pt]{<}}}}},
    degreeshift/.style={onearrow,dashed},
    octarine/.style={postaction={decorate,decoration={markings,mark=at position .6 with {\arrow[draw,line width=1pt]{>}}}}},
    tensor/.style={draw,double,double distance=1.5pt},
}

\newcommand\object[1]{\,\tikz[baseline=(basepoint)]\path[{#1}](0,-4pt) -- (0,12pt) (0,0) coordinate (basepoint);\,}

\newcommand\define[1]{{\em #1}}

\newcommand\Ff{{\mathcal F}}

\newcommand\Oo{{\mathcal O}}

\newcommand\NN{{\mathbb N}}
\newcommand\QQ{{\mathbb Q}}
\newcommand\RR{{\mathbb R}}
\renewcommand\SS{{\mathbb S}}

\newcommand\ZZ{\mathbb Z}

\newcommand\T{{\mathrm T}}
\renewcommand{\d}{{\mathrm d}}

\newcommand\C{{\mathrm C}}

\newcommand\q{{\mathrm q}}
\renewcommand\t{{\mathrm{inv}}}

\newcommand\st{{\textrm{ s.t.\ }}}

\DeclareMathOperator{\pexp}{pexp}

\DeclareMathOperator{\ave}{ave}

\DeclareMathOperator{\spec}{spec}
\DeclareMathOperator{\Hom}{Hom}
\DeclareMathOperator{\Sym}{Sym}
\DeclareMathOperator{\End}{End}

\DeclareMathOperator{\ev}{ev}
\DeclareMathOperator{\colim}{colim}

\DeclareMathOperator{\Chains}{Chains}

\DeclareMathOperator{\Homology}{H}
\renewcommand\H\Homology

\newcommand{\id}{\mathrm{id}}

\newcommand\cprime{\ensuremath'}

\DeclareMathOperator{\Frob}{Frob}
\DeclareMathOperator{\invFrob}{invFrob}
\DeclareMathOperator{\shFrob}{shFrob}
\DeclareMathOperator{\hFrob}{hFrob}
\DeclareMathOperator{\LB}{LB}
\DeclareMathOperator{\shLB}{shLB}
\DeclareMathOperator{\Pois}{PoisF}
\DeclareMathOperator\shBD{shBDF}
\newcommand\shBDF\shBD
\DeclareMathOperator\invLB{invLB}

\DeclareMathOperator{\qloc}{QLoc}

\newcommand\isom{\overset{\sim}{\to}}
\newcommand\mono{\hookrightarrow}
\newcommand\epi{\twoheadrightarrow}
\newcommand\onto\epi
\newcommand\longto{\longrightarrow}
\newcommand\<\langle
\renewcommand\>\rangle

\renewcommand\[{\llbracket}
\renewcommand\]{\rrbracket}

\newcommand\Circ{\ocircle}

\newcommand\shriek{{\text{\textexclamdown}}}

\title{Erratum in ``A combinatorial universal \texorpdfstring{$\star$}{star}-product''}
\author{Theo Johnson-Freyd}

\begin{document}

\maketitle

My paper ``A combinatorial universal $\star$-product'' \cite{rationalformal} contains an error rendering the central result incorrect.  I apologize to any reader who hoped to use that result, and thank Thomas Willwacher, without whose questions the error would not have been found.  The majority of that paper consists of correct ideas, which will appear in a new paper.  This note, which replaces \cite{rationalformal} on the arXiv, explains why the result is wrong, and where the error was in my proof.

\section{There is no wheel-free universal $\star$-product}

The central result claimed in \cite{rationalformal} was the existence of a wheel-free universal $\star$-quantization of Poisson infinitesimal manifolds.  Like any associative deformation problem, the problem of universal $\star$-quantization can be attacked directly, using a version of the Hochschild cochain complex.  For the question of wheel-free universal $\star$-products, there is a version whose $k$-cochains are universal wheel-free formulas for differential operators of $k$ inputs; the variable in these formulas is supposed to be a Poisson bivector.  The $\star$-product itself is a sequence $\{a_{n}\}_{n\in \NN}$ of $2$-cochains; thus the obstruction to extending a quantization to $n$th order is a $3$-cochain.  My claimed construction was such that $a_{n}$ scaled as $\lambda^{n}$ under rescaling the Poisson bivector by $\lambda$; I will call any cochain with this behavior ``$n$th-order.''

It is not too difficult to check by hand that there are no $2$-cocycles at $2$nd or $3$rd order.  It follows that, if there exists a wheel-free universal $\star$-product to $3$rd-order, then it is unique.  A $3$rd-order wheel-free universal $\star$-product was calculated by Penkava and Vanheacke \cite{MR1754236}, who also calculated the $4$th-order obstruction, and gave examples for which this obstruction is nonzero.  It follows that there is no wheel-free universal $\star$-product.

\section{Where my error lay}

My proposed construction of a universal $\star$-product involved building a (translation-invariant and ``quasilocal'') action on the (cellular) chains on $\RR$ of a cofibrant resolution of the properad $\Frob_{1}$ of $1$-shifted Frobenius algebras (i.e. the properad that controls the Frobenius algebra structure on $\H_{\bullet}(S^{1})$).  A more general result has been asserted by Wilson \cite{Wilson2007}
.  The space of such actions can be studied by obstruction theory.  Since $\Frob_{1} = \invFrob_{1}$ is Koszul, its minimal cofibrant resolution is easily calculated, and one can see directly that if an action exists (lifting the action on homology), then it is unique up to a contractible space of choices, and moreover that such an action exists provided finitely many obstructions vanish.  I proceeded to calculated these obstructions.

The error lies in the second display equation on page 10, wherein I gave an incorrect formula for the action of the generator $\left(   \,\tikz[baseline=(basepoint)]{ 
    \path (0,4pt) coordinate (basepoint) (0,2pt) node[dot] {} (-8pt,10pt) node[dot] {};
    \draw[](0,-6pt) -- (0,2pt);
    \draw[squiggly](0,2pt) -- (-8pt,10pt);
    \draw[](-8pt,10pt) -- (-12pt,18pt);
    \draw[](-8pt,10pt) -- (0pt,18pt);
    \draw[](0,2pt) -- (12pt,18pt);
  }\,
  -
  \,\tikz[baseline=(basepoint)]{ 
    \path (0,4pt) coordinate (basepoint) (0,2pt) node[dot] {} (8pt,10pt) node[dot] {};
    \draw[](0,-6pt) -- (0,2pt);
    \draw[squiggly](0,2pt) -- (8pt,10pt);
    \draw[](0,2pt) -- (-12pt,18pt);
    \draw[](8pt,10pt) -- (0pt,18pt);
    \draw[](8pt,10pt) -- (12pt,18pt);
  }\, \right)$, which is (one of) the syzygies weakening the coassociativity condition.  In the minimal resolution $\shFrob_{1}$ of $\Frob_{1}$, this generator transforms in a copy of the 2-dimensional irrep of $\SS_{3}$.  However, my formula transformed only in the permutation representation of $\SS_{3}$.  Thus my formula could not be correct; the correct formula should be the average of the one I gave under the appropriate $\SS_{3}$ action.  (The generators with two vertices imposing the Frobenius and counital axioms were correctly given.)
  
Using the correct formula, the obstruction to defining the action of the generator 
\begin{tikzpicture}[baseline=(basepoint)]
    \path (0,5pt) coordinate (basepoint) (0pt,2pt) node[dot] {} (8pt,12pt) node[dot] {} (-4pt,17pt) node[dot] {};
    \draw (0pt,-6pt) -- (0pt,2pt);
    \draw[squiggly] (0pt,0pt) -- (8pt,12pt);
    \draw[squiggly] (8pt,12pt) -- (-4pt,17pt);
    \draw[squiggly] (0pt,0pt) .. controls +(-6pt,6pt) and +(-6pt,-6pt) .. (-4pt,17pt);
    \draw (8pt,12pt) -- (12pt,24pt);
    \draw (-4pt,17pt) -- (-4pt,24pt);
  \end{tikzpicture}
does not vanish at the chain level, although it does vanish at homology.  In particular, this generator cannot be taken to act as $0$.  There is one final generator, this one with genus $2$ and four vertices: 
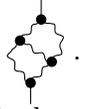
\begin{tikzpicture}[baseline=(basepoint)]
    \path (0,12pt) coordinate (basepoint) (0pt,3pt) node[dot] {} (8pt,11pt) node[dot] {} (-4pt,19pt) node[dot] {} (4pt,27pt) node[dot]{};
    \draw (0pt,-5pt) -- (0pt,3pt);
    \draw[squiggly] (0pt,3pt) -- (8pt,11pt);
    \draw[squiggly] (8pt,11pt) -- (-4pt,19pt);
    \draw[squiggly] (0pt,3pt) .. controls +(-6pt,6pt) and +(-6pt,-6pt) .. (-4pt,19pt);
    \draw[squiggly] (-4pt,19pt) -- (4pt,27pt);
    \draw[squiggly] (8pt,11pt) .. controls +(6pt,6pt) and +(6pt,-6pt) .. (4pt,27pt);
    \draw (4pt,27pt) -- (4pt,35pt);
  \end{tikzpicture}\,.  Upon calculating its obstruction, one arrives at a non-zero multiple of the identity, and this does not vanish in homology.  This is almost exactly the obstruction found by Merkulov \cite{Merkulov07}.

The end result of these calculations is, I think, quite surprising.  There does not exist a translation-invariant and quasilocal action on $\Chains_{\bullet}(\RR)$ of any cofibrant replacement of the properad $\Frob_{1}$, at least not one that induces both the comultiplication on homology and the multiplication on cohomology.  Translation invariance and quasilocality are very natural conditions to request if the chain-level structure is to have a geometric interpretation.

%

\end{document}